\documentclass[manuscript,screen]{acmart}

\AtBeginDocument{%
  }

\setcopyright{acmlicensed}
\copyrightyear{2024}
\acmYear{2024}
\acmDOI{XXXXXXX.XXXXXXX}
\usepackage{placeins}
\usepackage{array}
\usepackage{float}

\begin{document}

\title{Towards \textit{in-situ} Psychological Profiling of Cybercriminals Using Dynamically Generated Deception Environments}
\author{Jacob Quibell}
\email{jacob.quibell@gmail.com}
\orcid{0000-0001-9830-9839}
\renewcommand{\shortauthors}{Quibell}
\begin{abstract}
  Cybercrime is estimated to cost the global economy almost \textdollar 10 trillion annually and with businesses and governments reporting an ever-increasing number of successful cyber-attacks there is a growing demand to rethink the strategy towards cyber security. The traditional, perimeter security approach to cyber defence has so far proved inadequate to combat the growing threat of cybercrime.  Cyber deception offers a promising alternative by creating a dynamic defence environment. Deceptive techniques aim to mislead attackers, diverting them from critical assets whilst simultaneously gathering cyber threat intelligence on the threat actor. This article presents a proof-of-concept (POC) cyber deception system that has been developed to capture the profile of an attacker \textit{in-situ}, during a simulated cyber-attack in real time.  By dynamically and autonomously generating deception material based on the observed attacker behaviour and analysing how the attacker interacts with the deception material; the system outputs a prediction on the attacker’s motive. The article also explores how this POC can be expanded to infer other features of the attacker’s profile such as psychological characteristics.document cl
\end{abstract}

\received{}
\received[revised]{}
\received[accepted]{}
\maketitle
\section{Introduction}
Cyber Deception is emerging as a promising defensive strategy that can augment traditional solutions \cite{Zhang2021Three}. Instead of targeting the attacker’s actions, Cyber Deception targets their perception. By confusing or misleading attackers using decoy assets and information, attacks can be delayed and disrupted, increasing the cost to the adversary. Cyber Deception also offers an opportunity to engage with the attacker and manipulate activity in order to gain unique outcomes such as long-term deterrence and collection of precise threat intelligence \cite{Ferguson-Walter2021Examining}. This has also opened up the cyber defence domain to other lines of scientific study, particularly in the field of psychology, \cite{Ferguson-Walter2021Oppositional} where studies investigating the application of human factors \cite{Ferguson-Walter2021Oppositional} and game theory \cite{Ferguson-Walter2019Game} show the rich, multidisciplinary nature of Cyber Deception and the potential for knowledge in other domains to be applied to cyber security.

Typically, Cyber Deception campaigns deploy static deception assets such as honeypots or honeytokens, which can serve as alarms to intrusions or distract attackers, consuming time and resources \cite{Rowe2016Introduction}.  While these techniques can be useful to disrupt cyber-attacks, Cyber Deception can perform potentially more valuable activities such as threat intelligence collection \cite{Cho2020Toward}. Cyber threat intelligence has emerged over the past few years to help security practitioners  recognise the indicators of cyber-attacks, extract information about the attack methods, and consequently responding to the attack with greater accuracy and speed \cite{Conti2018Cyber}. 

This work presents a Proof-of-Concept (POC) system that employs a novel cyber deception technique to capture cyber threat intelligence. The POC has been built and deployed on cloud infrastructure and basic testing has been performed successfully. The impacts of these results and the potential avenues for future research is also discussed. 

The rest of the article is organised as follows: In Section \ref{relatedWork} the current methods for profiling cybercriminals are evaluated. In Section \ref{approach} the approach to achieving the research goals is presented, including the attack scenario, system design, architecture and workflow. In Section \ref{discussion} the outcomes of the work will be discussed along with the constraints and future work before concluding in Section \ref{conclusion}. Appendices with supporting figures and tables are found at the end of the article.

\section{Related Work} \label{relatedWork}

There exists a range of techniques to collect threat intelligence, but it is generally done by analysing large datasets of previous cyber-attacks or open-source information on the internet to detect attack approaches, behaviours and patterns \cite{Sun2023Cyber}. Deception environments, or honeypots, have also been used to capture raw data to then process into threat intelligence, but again the threat intelligence is only developed after the attack has occurred on the deception environment \cite{Almohannadi2018Cyber, SokolData}. There has never been an attempt to collect this data using deception \textit{in-situ}, whilst the attack is ongoing, and crucially, leverage that information to achieve the goals of the live deception campaign. 

A key element of the threat intelligence picture is the criminal profile. This is a key tool to investigators and is used to narrow the range of suspects and evaluate the likelihood of a suspect committing a crime. It consists of a set of characteristics likely to be shared by criminals who commit a particular type of crime and combines: personal traits of the cybercriminal, behavioural patterns, demographic data, motivations and psychological traits \cite{Martineau2023Comprehensive, Rogers2016Chapter}. It is typically used in criminal investigations for attribution but can have impact on improving network defence by developing a more mature threat model, and therefore implementing more targeted defences based on skills/motivations of the threat actor.

There are many methods to profile a cybercriminal which approach the task from a variety of angles, however one key factor is common to them all; they are all conducted after the attack has taken place \cite{Bada2021Profiling}. This is understandable given that the traditional approach to cyber incident response eradicates the threat before any forensic activities take place \cite{Lickiewicz2011Cyber}.

Another gap in the literature is the extent to which psychological data is captured in the cybercriminal profile. Psychological profiling is of importance in criminal investigations to determine a relationship between the attacker’s personality and the crime committed. This leads to an inference of the modus operandi – a key piece of information used in law enforcement to attribute crimes to individuals or groups \cite{Rogers2016Chapter}. There have been a number of articles describing the importance of this and have proposed frameworks and methodologies demonstrating how psychological profiling can be conducted using forensics collected from the scene of the cybercrime \cite{Lickiewicz2011Cyber, Chng2022Hacker}, however few collected primary data and none use data gathered from deception environments as the primary data source \cite{Bada2021Profiling}. Usually, this data is gathered through interviews, questionnaires and other psychometric tests with known cyber offenders; however, as cyber criminals are notoriously difficult to arrest, there is a distinct lack of opportunity to collect psychological data through these means \cite{Rogers2006Self-reported}. 

There have been several studies investigating how to programmatically profile an attacker based on the behaviour captured across a network. Examples of these include using methods such as Fuzzy Inference \cite{Mallikarjunan2018Real}, hidden Markov models \cite{Katipally2011Attacker} and attack graph analysis \cite{Casey2007Threat} to infer characteristics of the attacker, and demonstrate that there is a burgeoning body of knowledge in how different mathematical models can be applied to various datasets to achieve reliable predictions on the attacker’s profile. Honeypots have also been used to capture threat actor behaviour which is then analysed to infer the profile \cite{Fraunholz2017YAAS}. The results of these studies are promising, however they all use static captures of the malicious behaviour and do not respond to the threat in real time.
\section{Approach} \label{approach}
Given the lack of opportunity to profile cybercriminals by traditional means, it would therefore be beneficial if this information could be captured at the most critical opportunity to interact with attackers – during a live attack. This proof-of-concept looks to explore the possibility of this \textit{in-situ} data collection by delivering targeted deception material to the attacker, the way in which the attacker interacts with this material, elucidates the attacker's profile. This real-time behavioural analysis could provide a unique and highly valuable method to collect intelligence on attackers. 

A crucial addition to the methodology proposed here is the dynamic generation of deceptive content based on the information captured throughout the attack. The system will then automatically deploy the new deceptive content to a new environment and the process iterates (Figure 1). By iterating over several deception environments, the intelligence gained is refined and reinforced. This iterative, dynamic and automated approach to in-situ threat intelligence collection is a core concept of this research and currently unexplored in previous academic work. There may exist the possibility to collect richer psychological data with this method such as biases, personality traits, or even affective states such as confusion or self-doubt; this is discussed in Section \ref{futureWork}. However, for the purposes of this proof-of-concept (POC), the scope of the criminal profile is restricted to motive \cite{Martineau2023Comprehensive}.
\begin{figure}
    \centering
    \includegraphics[width=0.75\linewidth]{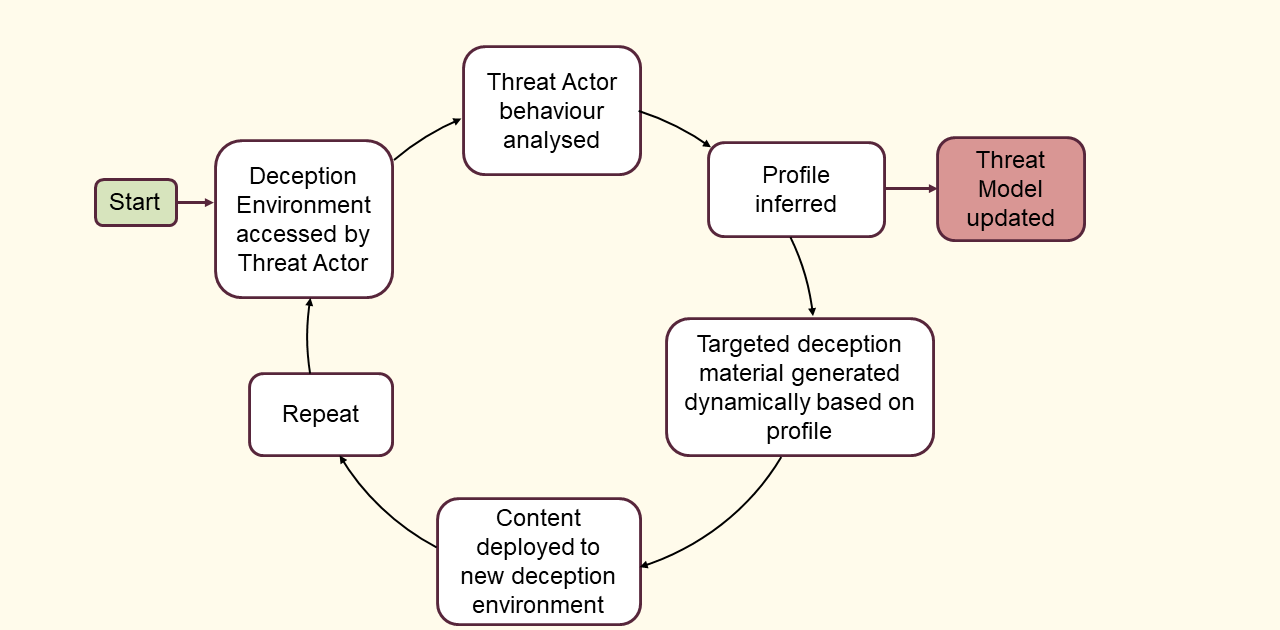}
    \caption{High-level workflow of prediction system}
    \label{fig:enter-label}
\end{figure}
\subsection{Attack Scenario}
To begin developing this deceptive system that reacts to malicious activity, the scope was narrowed to target a specific attacker behaviour. To do this, a particular attack scenario was selected to drive the design of the simulated environment. The scenario proceeds as such: the attacker has gained access to a corporate network and is moving laterally through the system to conduct discovery and/or collection activities. During this lateral movement, the attacker accesses a subnet consisting of a set of fileshare servers, which host a variety of different word documents containing different types of information on the victim organisation. Specifically, the types of information held on these fileshares are Financial, Operational, Human Resources (HR), Legal and Information Technology (IT). 

The attacker then initiates a systematic exploration of the fileshares, examining each one in sequence, aiming to identify information within the word documents pertinent to their underlying motive. In this attack scenario, the subnet accessed by the attacker is a deception environment where the fileshare hosts are instrumented, so that all attacker activity on the host is captured without the knowledge of the attacker. It is when the attacker accesses this environment and begins exploring the Word documents that the predictive system is activated.

\subsection{System Design} \label{SystemDesign}

The objective of the system is to generate a prediction of the attacker motive based on the type of information that is accessed. To enable this inference, a relationship between the types of documents present on the system and threat actor motives must first be established. These relationships are mapped in Table \ref{tab:table1}. One of the constraints in the system design of this POC is the 1-to-1 relationship between motive and document type. A more rigorous and realistic approach would be to establish a 1-to-many relationship between motive and document type with weightings to reflect the relative value of each document type to a threat actor with a particular motive. Due to a lack of evidence in the literature describing such relationships, and to enable the timely development of the POC, the mapping between motive and document type was chosen to be a 1-to-1 relationship determined by intuitive use cases. Examples of these are described in Table \ref{tab:table1}.

\begin{table*} 
  \renewcommand{\arraystretch}{1.5}
  \caption{Mapping of Motive to Document Type with explanatory Use Case}
  \label{tab:table1}
  \begin{tabular}{ccp{10cm}}
    \toprule
    Motive&Document Type&Use Case\\
    \midrule
    Profit & Financial& The attacker is part of a criminal organisation specialising in ransomware. The group’s motive is generating profit and so are particularly interested in financial documents\\
    Ideological & HR & The attacker is a hacktivist and is ideologically motivated. They want to expose the suspected immoral practices of the victim organization and the individuals behind them. The attacker is therefore interested in HR documents that reveal employee details.\\
    Geopolitical & Operational & The attacker is part of an APT group backed by a nation state who are geopolitically motivated. The group is instructed to steal Intellectual Property from the victim organisation and so are interested in Operational documents.\\
    Satisfaction & IT & The attacker is lone thrill seeker and is motivated by satisfaction. The attacker’s goal is to compromise the most secure server on the network and so is interested in IT documentation.\\
    Discontent & Legal & The attacker is a disgruntled employee presenting an insider threat and wants to cause harm to the victim organisation's reputation. The attacker is therefore interested in any Legal documents that detail embarrassing disputes the organisation was involved in.\\
  \bottomrule
\end{tabular}
\end{table*}

An initial set of deception documents was then generated, the subjects for each document type is shown in Appendix A. The text for these documents was generated by OpenAI’s gpt-3.5-turbo model. To maximize the authenticity of the document the model was asked to produce text that resembled files from the Pandora Papers leak in 2021 \cite{2021Pandora}. To generate the desired text the guardrails required circumventing by asking the model to behave like a ‘movie prop text writer’. The prompt can be found in Appendix B. More convincing text can undoubtedly be generated by populating the prompt with greater detail on the background of the victim organization but for the purposes of this POC, these results were sufficient. An example document can be found in Appendix C. The files were given generic names according to the subject they belonged to (e.g. IT Asset Inventory01) and then downloaded to a single directory on the first fileshare: Deception Env 1.

When a document is generated a set of attributes is automatically assigned to it. These attributes are described in Table \ref{tab:table2}. The attributes are stored in an external database that can be queried when a particular document is accessed in the deception environment. Thus, the system can identify which documents have been accessed and what motive a particular document relates to.

\begin{table*}
  \renewcommand{\arraystretch}{1.5}
  \caption{Document Metadata}
  \label{tab:table2}
  \begin{tabular}{ccp{10cm}}
    \toprule
    Attribute&Type&Description\\
    \midrule
    locHash & String & Sha256 hash of the document location that provide unique identifier for the document. Location described in JSON format with the absolute path on the host and the host name i.e. 
    \begin{verbatim}
    {
        ‘path’: absolute_path,
        ‘host’: hostname
    }\end{verbatim}\\
    deception\_host & Integer & Identifier for the deception host the document is deployed to
    \\
    motives & JSON Object & Identifies which motive the document is related to (e.g. ideological, discontent, geopolitical, satisfaction, profit)\\
    subject & String & The subject the document relates to (e.g. Asset Inventory) \\
    type & String & The type of document produced (e.g. Legal, Financial, HR, Operations or IT) \\
  \bottomrule
\end{tabular}
\end{table*}

\subsection{System Architecture}

\begin{figure}
    \centering
    \includegraphics[width=0.75\linewidth]{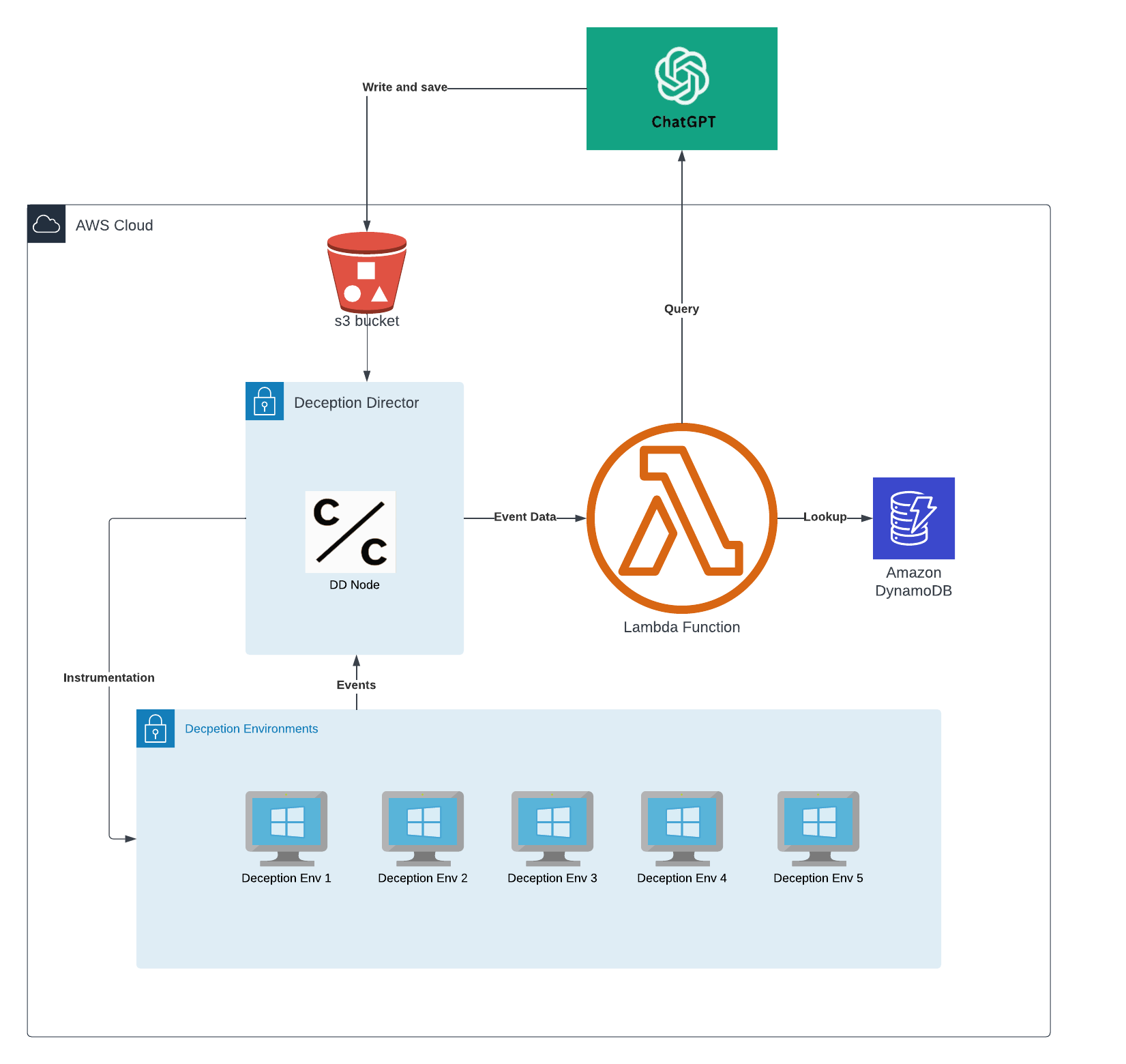}
    \caption{System Architecture}
    \label{fig:Arch}
\end{figure}

The infrastructure on which this system was deployed was Amazon Web Services (AWS), with the exception of OpenAI’s ChatGPT API service, which was accessed over the internet. Figure \ref{fig:Arch} depicts the architecture of the system with the individual components summarized in Table \ref{tab:table3}. The system leverages proprietary software from CounterCraft © to instrument the deception environments. This selection was due to the availability and familiarity of the software. Other monitoring tools (e.g. ELK, Splunk) could easily be substituted to perform the same function.

\begin{table*}
  \renewcommand{\arraystretch}{1.5}
  \caption{System Components}
  \label{tab:table3}
  \begin{tabular}{>{\centering\arraybackslash}p{2.5cm}>{\centering\arraybackslash}p{2.5cm}>{\arraybackslash}p{9cm}}
    \toprule
    Component&Deployment&Function\\
    \midrule
    Deception Environments & AWS EC2 
    (Windows server) & Hosting deception material\\
    Deception Director 
    (CounterCraft software) & AWS EC2 
    (Ubuntu server) & Instrumentation of deception environments. Monitoring for specific events. Hosting logic to send SNS message to Lambda function containing event data. Hosting logic to download deception material to deception environments.\\
    Lambda Function & AWS Lambda & Hosting logic to analyse threat actor behaviour and perform programmatic profiling of threat actor\\
    DynamoDB & AWS DynamoDB & Storage of deception material attributes \\
    ChatGPT API & OpenAI Servers & Generation of deceptive text \\
    S3 Bucket & AWS S3 & Storage of deceptive files\\
  \bottomrule
\end{tabular}
\end{table*}

\subsection{Workflow}

\textit{Phase 1: Monitor}

As mentioned, the fileshares are instrumented so that every event on the deception host is monitored by the Deception Director. A rule is then deployed in the Deception Director which sends a SNS message to trigger a Lambda function, any time a word document is opened on the deception host. The details of the event including the absolute path of the file opened is contained in the SNS message.

 \noindent \textit{Phase 2: Behaviour Analysis}

The lambda function then performs a series of steps to identify which motive is assigned to that document. A score is given to the motive by examining the sequence in which the documents are accessed in a given attack on a particular deception environment (e.g. 1st document accessed in deception environment – motive score = 100, 2nd document accessed in deception environment – motive score = 80, etc.) For this POC, the attack on the deception environment is set to be complete when 6 documents are accessed. This was an arbitrary decision and the number of documents accessed before the next phase of the workflow is triggered can be easily reconfigured. Once the 6th document is accessed in the deception environment the motive scores are aggregated and ranked. An example of the scoring is shown in Table \ref{tab:scoring} with the final aggregated scores shown in Table \ref{tab:aggscores}.
\begin{table*}
  \renewcommand{\arraystretch}{1.5}
  \caption{Example Scoring}
  \label{tab:scoring}
  \begin{tabular}{ccccc}
    \toprule
    Position&Document Subject&Document Type&Associated Motive&Score\\
    \midrule
    1/6&Annual Budget&Financial&Profit&100\\
    2/6&IT Asset Inventory&IT&Satisfaction&80\\
    3/6&Standard Operating Procedures&Operational&Geopolitical&60\\
    4/6&Corporate Governance Documents&Legal&Discontent&40\\
    5/6&Tax Document&Financial&Profit&20\\
    6/6&Employment Contracts&HR&Ideological&0\\
  \bottomrule
\end{tabular}
\end{table*}

\begin{table*}
  \renewcommand{\arraystretch}{1.5}
  \caption{Example Aggregated Scores}
  \label{tab:aggscores}
  \begin{tabular}{cc}
    \toprule
    Motive&Aggregated Score\\
    \midrule
    Profit&120\\
    Satisfaction&80\\
    Geopolitical&60\\
    Discontent&40\\
    Ideological&0\\
  \bottomrule
\end{tabular}
\end{table*}

 \noindent \textit{Phase 3: Motive Elimination and Document Generation}

In this phase, the system ranks the motives according to their score and eliminates the lowest scoring motive from the motive list. This revised list is then used to generate a new set of documents in the same way as described in Section \ref{SystemDesign}. This set of documents will not contain any document types associated with the lowest scoring motive. In the example given above this would be HR documents associated with the Ideological motive. These newly generated documents will then be deployed to the subsequent deception environment.

 \noindent \textit{Phase 4: Iteration}
 
The previous steps now repeat for the remaining deception environments. After each iteration is complete a motive is eliminated from the motive list and new documents are deployed. The workflow ends when only one motive remains. This is the system's final prediction of the attackers motive.
\begin{figure}
    \centering
    \includegraphics[width=1\linewidth]{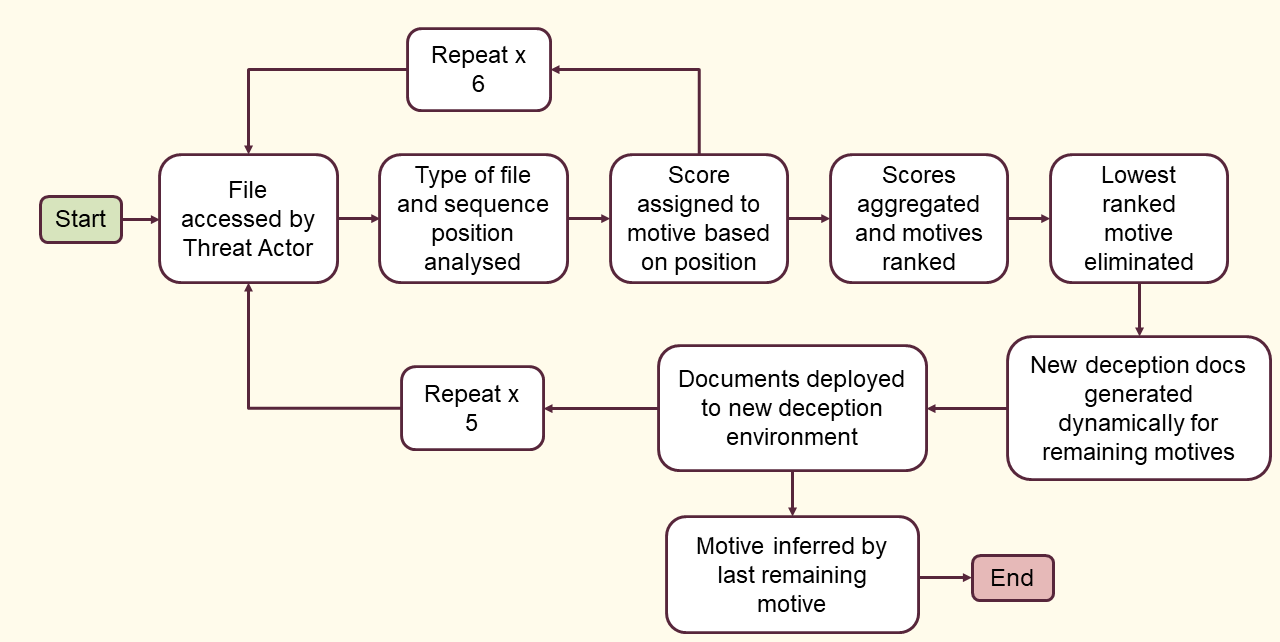}
    \caption{Prediction function workflow}
    \label{fig:figure3}
\end{figure}

\subsection{Testing}

Informal testing of the system was undertaken to investigate the basic functionality. In this activity five individuals were asked to read a selected use case from Table 1 and play the role of the attacker with the profile described in the use case. The participants were then set the task of exploring the filshares for information pertinent to the goals of attacker role they assumed. This activity was supervised to ensure the participants proceeded through the exercise in the way that it was intended. Within a particular fileshare, the participants were allowed to access the files in any order. Once they had completed exploring a fileshare, they were guided by the supervisor to the next deception environment where new deception material had just been deployed. This continues until all deception environments had been explored. Pleasingly the system predicted the correct motive in every exercise. This result is not indicative of the system's potential success in realistic attack scenarios, but does demonstrate that the POC functions as intended in dynamically generating targeted deception material and providing an inference on attacker motive.

\section{Discussion} \label{discussion}
\subsection{Outcomes}
This POC has demonstrated a method to map low-level attacker behaviour (file access) to higher level attacker characteristics (motive). This is a challenging task even with advanced machine learning analytics \cite{Kotenko2023Cyber}. The novel approach presented in this paper - analysing attacker behaviour over several iterations, enabled by deception, has been shown to be a promising method for maximising the probability of an accurate inference of higher-level attacker characteristics. Additionally, the system is entirely automated removing the need for human interaction and thus allows for a response to occur in real time, as the attack proceeds.

\subsection{Constraints}
This POC system is highly constrained and not reflective of a realistic attack. The constraints were put on the system to allow for initial development and should the system continue development to a production ready application, these constraints will need to be addressed. The most significant constraints identified are described below.

This deceptive system currently only targets a very narrow range of the cyber kill chain \cite{Hutchins2011Intelligence-driven}, namely the Actions on Objective or the Collection Phase in the MITRE ATT\&CK framework \cite{MITRE}. At this point the attacker has conducted most of their activities and usually about to exfiltrate the selected data and end the attack. To be an effective countermeasure against cyber-attacks the scope should be expanded to disrupt earlier stages of the kill chain. 

In addition, this POC relies on specific attacker behaviour i.e. moving from one deception environment to the next in sequence. In reality, the way in which attackers access environment may be unpredictable, this means that the deception campaign must be carefully planned to maximise the chances of the attacker behaving in the intended way. One method for this would be to plant breadcrumbs in initial deception environments that contain information (e.g. credentials) that leads the attacker to a desired deception environment. Other challenges, such as the latency associated with generating and deploying the deception material and the authenticity of the deception material also need to be investigated.

Finally, the predictive function used to infer the motive behind the attack is a rudimentary process of elimination method. This will require a significant revision to reflect the much more complicated relationship between the types of information accessed by the attacker and the underlying motive. Approaches to addressing these limitations are discussed in more detail in the following section.

\section{Future Work} \label{futureWork}
Given this system is a POC, there exist many avenues for further development. In this section, the core areas of future research are explored. While this is not an exhaustive list, completion of the following work would provide a much clearer perspective on which direction to focus future research.

\subsection{Expansion of system to include network services}
As described, the POC only examines attacker interaction with Word documents. For any utility in a real attack, the system must contain deceptive versions of services and assets that are expected in an enterprise network. This may include deceptive email and web services, databases, workstations and active directories, all of which need to be monitored for activity. Importantly, these services need be deployed dynamically, and populated with deceptive content based on the observed attacker behaviour. As such, an API should be developed with the capability to facilitate this dynamic deployment. As with the research presented earlier in this article, the deceptive content could be generated by ChatGPT.

\subsection{System validation and data harvesting}
To fully validate the effectiveness of the deception strategy, exercises must be carried out with human participants playing the role of the attacker. The participants should not be aware of the deceptive nature of the system and explore the system for information pertinent to the motive of that threat actor. This could reveal which actions the attacker is likely to take and in what order, providing insights into how to manipulate attacker behaviour to ensure the desired outcomes are reached.

As with the POC, a matured system with deceptive network services would react to the behaviour of the participants and deploy targeted deception assets in continually evolving deception environments.
The primary goal of these exercises would be to generate data to inform further development. Thus, the entirety of the participant behaviour across the network throughout the exercise should be captured. This dataset would provide a clear link between malicious activity on a network and the threat actor motive.

\subsection{Development of model for behaviour analysis}
Undoubtedly the predictive power of this system would be significantly enhanced by the application of a machine learning model. The data captured in the validation phase could be used to train this model which could then output predictions of the attacker motive.

The use of neural networks to profile attackers based on network behaviour has been attempted before by Kotenko et. al \cite{Kotenko2023Cyber}. This research only yielded moderate results due to the lack of appropriate datasets; however, it demonstrates the feasibility of a machine learning approach. Future work here could look to use similar techniques as described by Kotenko but would benefit from having a labelled dataset of network behaviour generated specifically for training the model. This should maximise the potential performance of any model developed, and deliver a robust prediction on the profile on the attacker. Additionally, this prediction can be reinforced as the attacker iterates over multiple deception environments.

\subsection{Psychological profiling}
The most impactful and interesting application of this system would be the investigation into how psychological features of the attacker’s profile can be inferred though the dynamic deployment of deceptive content and the attacker’s interaction with it. To enable this investigation of psychological profiling, the capture of psychological information at certain points in the data gathering exercises described above should be carried out. Capturing psychological information when then participant is exposed to a particular deceptive artefact will provide an association between their cognitive state and the network behaviour exhibited on the system. Capturing this information through Cyber Task Questionnaires has been demonstrated before in the literature \cite{Ferguson-Walter2021Oppositional}. Questionnaires can also be completed after the exercises to capture the experiences of the participants \cite{Gutzwiller2019Are}. The aim is to map psychological states and/or biases to network behaviour to so that a network behaviour dataset, labelled with psychological information can be developed. These datasets could then be leveraged in the development of a model to output predictions on the psychological profile of the attacker. 

Some exploratory investigations into the future work described here have been conducted, but much of the research remains aspirational. The question of whether attackers are susceptible to decision making biases, and does the interaction with specific deception artefacts reveal these cognitive biases, remains unanswered. Answering this fascinating question requires a multidisciplinary approach and must combine the knowledge in behavioural science with programming and data science techniques. 

\section{Conclusion} \label{conclusion}
A fully automated and responsive cyber deception system has been developed. This system analyses attacker behaviour and generates deception material based on what is observed. Through iterating this activity over several deception environments, the system can produce a refined prediction of the underlying motive of the attacker.

At its most fundamental level, what the system presents is a communicative interface with the attacker. The selection of deceptive material by the attacker is comparable to a multiple-choice questionnaire that with each selection reveals more about the human behind the attack. This communication with the attackers could open the door to a whole raft of potential experiments, bringing in knowledge and techniques from psychology, criminology and game theory. 

The challenges with implementing this research are not underestimated, validating the effectiveness of this deception technique would need careful experimental design. In addition, how this highly constrained system could be translated to real world networks would require significant effort. Nevertheless it is the author’s hope that the work described, and the proposals contained in the article will provide a rich ground for discussion in attempting to move beyond purely technical responses to attacks and usher in a more psychological focus to cyber security.
\begin{acks}
I would like to thank my colleague Darren Lawrence for his invaluable support and guidance. I would also like to thank the CounterCraft team for providing a license to their software free of charge for this research. Finally, I would like to thank my wife Dejana for her unwavering love and support.
\end{acks}
\bibliographystyle{ACM-Reference-Format}
\bibliography{bibliography}
\clearpage
\appendix
\section{Appendix A: Subjects of document types used}
\begin{table*}[hb]
  \renewcommand{\arraystretch}{1.5}
  \caption{Example scoring}
  \label{tab:table5}
  \begin{tabular}{>{\centering\arraybackslash}p{2.5cm}>{\centering\arraybackslash}p{2.5cm}>{\centering\arraybackslash}p{2.5cm}>{\centering\arraybackslash}p{2.5cm}>{\centering\arraybackslash}p{2.5cm}}
    \toprule
    Financial & HR & IT & Legal & Operational \\
    \midrule
    General Ledger & Time and Attendance Records & IT Asset Inventory & Non-Disclosure Agreements (NDAs) & Safety Procedures \\

    Tax Documents & Employee Benefit Documents & IT Policies and Procedures & Compliance Documentation & Standard Operating Procedures (SOPs) \\
    
    Financial Contracts & Training and Development Plans & Security Policies and Procedures & Corporate Governance Documents & Change Request Forms \\
    
    Payroll Documents & Employee Handbook & Vendor Contracts and Service Level Agreements & Litigation and Legal Proceedings Documents & Inventory and Stock Control Documents \\
    
    Compliance and Regulatory Documents & Employee Records & Disaster Recovery and Business Continuity Plans & Legal Opinions and Memoranda & Incident Reports \\
    
    Budgets & Exit Interview Forms & System Documentation & Policies and Procedures & Performance Metrics and Dashboards \\
    
    Financial Statements & Performance Appraisal Forms & Change Management Documents & Regulatory Filings & Maintenance and Equipment Manuals \\
    
    Financial Reports & Offer Letters & IT Project Documentation & Legal Research and Case Law & Quality Control Documents \\
    
    Audited Financial Statements & Employment Contracts & Incident and Problem Reports & Contracts & Work Instructions \\
    
    Invoices and Purchase Orders & Job Descriptions & IT Service Level Agreements (SLAs) & Intellectual Property Documents & Production Plans \\
  \bottomrule
\end{tabular}
\end{table*}
\clearpage
\section{Appendix B: ChatGPT query written in Python with document type passed in as variable}
\begin{verbatim}
    {"role": "user", "content":
        """
            You are a world class movie prop text writer.
            I need you to draft a financial document that looks EXACTLY like a document 
            that would look like it was included in the corpus of financial documents recovered
            in the Pandora Papers scandal for a dramatized version of the event. It must be as 
            accurate as possible for realism. It should be for a company called Jacob & Co Ltd,
            a hedge fund based in Gibraltar and Panama. The Type of document should be a
            """ + str(document_type) + """
            Respond with the the document in a well-organized text file.
            Think through it carefully before responding. 
            Only respond with the document directly, do not add any additional
            commentary or explanation (for realism).
        """
    }
\end{verbatim}
\clearpage
\section{Appendix C: Example financial document generated by the system}
\begin{verbatim}
Title: General Ledger of Jacob & Co Ltd
Period: January 2021 to December 2021
Account: Cash
---------------------------------------
Date         Description                  Debit      Credit      Balance
------------------------------------------------------------------------
01/01/2021   Opening Balance                    -            -     5,200,000
01/15/2021   Investment Received      10,000,000            -    15,200,000
02/07/2021   Transfer to Invest. Acct           -   12,000,000     3,200,000
07/31/2021   Hedge Fund Returns        14,400,000            -    17,600,000
12/31/2021   Operating Expenses                -     1,200,000    16,o00,000
Account: Investment Account
---------------------------------------
Date         Description                  Debit      Credit      Balance
------------------------------------------------------------------------
01/01/2021   Opening Balance                    -            -    25,000,000
02/07/2021   Transfer from Cash Acct  12,000,000            -    37,000,000
07/31/2021   Hedge Fund Investments          -     14,400,000    22,600,000
Account: Capital
---------------------------------------
Date         Description                  Debit      Credit      Balance
------------------------------------------------------------------------
01/01/2021   Opening Balance                    -            -    50,000,000
01/15/2021   Capital Invest. Received          -    10,000,000    60,000,000
Account: Income
---------------------------------------
Date         Description                  Debit      Credit      Balance
------------------------------------------------------------------------
01/01/2021   Opening Balance                    -            -             0
07/31/2021   Hedge Fund Income                 -    14,400,000    14,400,000
Account: Expenses
---------------------------------------
Date         Description                  Debit      Credit      Balance
------------------------------------------------------------------------
01/01/2021   Opening Balance                    -            -             0
12/31/2021   Operating Expenses        1,200,000            -     1,200,000
=-=-=-=-=-=-=-=-=-=
End of General Ledger
=-=-=-=-=-=-=-=-=-=
\end{verbatim}
\clearpage
\section{Appendix D: Model of end-to-end workflow}
\begin{figure}[hb]
    \centering
    \includegraphics[width=1\linewidth]{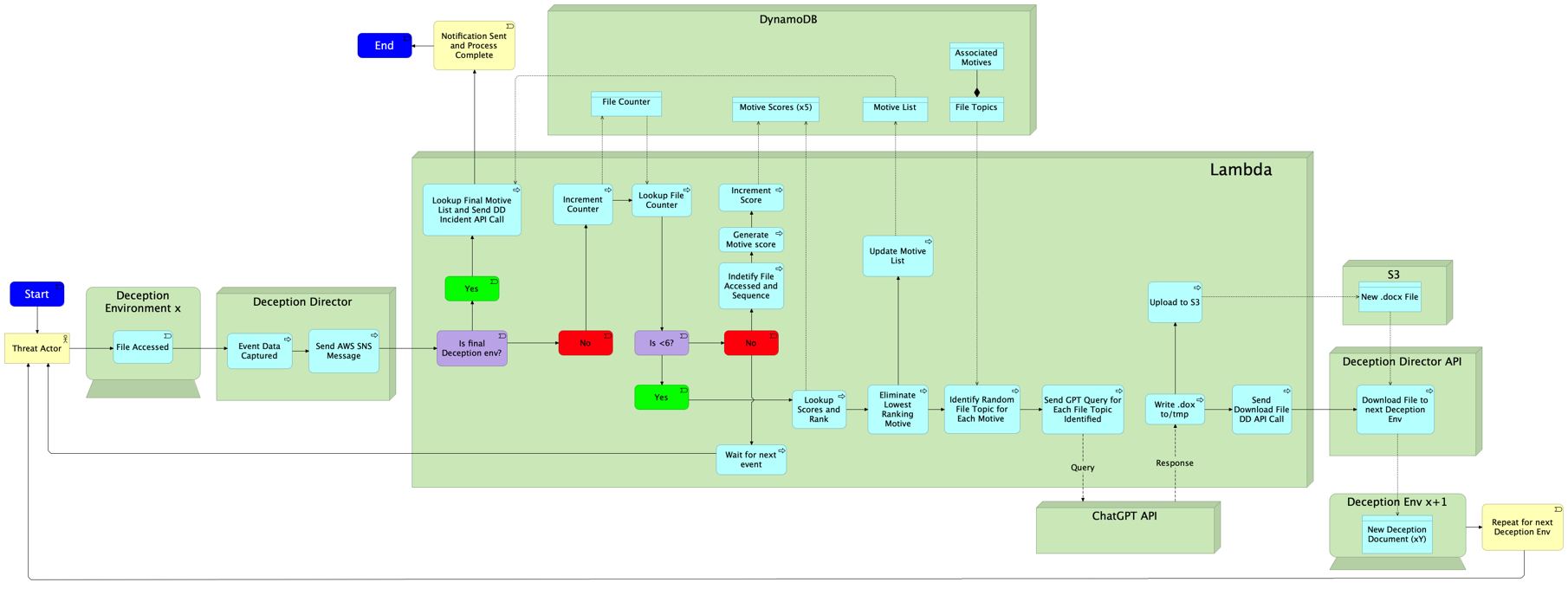}
    \caption{Model of end-to-end workflow}
    \label{fig:model}
\end{figure}
\end{document}